\documentclass{article}

\usepackage{amssymb,amsfonts,amsmath}
\usepackage{cite,enumerate,float}
\usepackage{color}
\usepackage{tikz}
\usetikzlibrary{arrows,snakes,backgrounds}
\usepackage{graphicx}
\usepackage{amsthm}
\usepackage{framed}
\usepackage{mathtools}

\def\be{\begin{eqnarray}}
\def\ee{\end{eqnarray}}
\def\nn{\nonumber}

\def\p{\partial}

\definecolor{red}{rgb}{1,0,0}
\definecolor{orange}{rgb}{1,0.5,0}
\definecolor{violet}{rgb}{0.7,0,1}

\def\S{\mathfrak{S}}
\def\NS{\mathfrak{NS}}

\usepackage[cmtip,all]{xy}

\newcommand{\longsquiggly}{\xymatrix{{}\ar@{~>}[r]&{}}}

\hoffset= - 1.0in         

\begin{document}

\title{\vspace{1.5cm}\bf
Diamond of triads
}

\author{
A. Mironov$^{b,c,d,}$\footnote{mironov@lpi.ru,mironov@itep.ru},
A. Morozov$^{a,c,d,}$\footnote{morozov@itep.ru},
A. Popolitov$^{a,c,d,}$\footnote{popolit@gmail.com},
Z. Zakirova$^{e,d}$\footnote{zolya\_zakirova@mail.ru}
}

\date{ }

\maketitle

\vspace{-6cm}

\begin{center}
\hfill MIPT/TH-04/25\\
\hfill  FIAN/TD-03/25\\
\hfill ITEP/TH-04/25 \\
\hfill IITP/TH-04/25
\end{center}

\vspace{3cm}

\begin{center}
$^a$ {\small {\it MIPT, Dolgoprudny, 141701, Russia}}\\
$^b$ {\small {\it Lebedev Physics Institute, Moscow 119991, Russia}}\\
$^c$ {\small {\it NRC ``Kurchatov Institute", 123182, Moscow, Russia}}\\
$^d$ {\small {\it Institute for Information Transmission Problems, Moscow 127994, Russia}}\\
$^e$ {\small {\it Kazan State Power Engineering University, Kazan, 420066, Russia}}
\end{center}

\vspace{.1cm}

\begin{abstract}
The triad refers to embedding of two systems of polynomials, symmetric ones and those of the Baker-Akhiezer type into a power series of the Noumi-Shiraishi type. It provides an alternative definition of Macdonald theory and its extensions. The basic triad is associated with the vector representation of the Ding-Iohara-Miki (DIM) algebra.
We discuss lifting this triad to two elliptic generalizations and further to the bi-elliptic triad. At the algebraic level, it corresponds to elliptic and bi-elliptic DIM algebras. This completes the list of polynomials associated with Seiberg-Witten theory with adjoint matter in various dimensions.
\end{abstract}

\bigskip

\newcommand\smallpar[1]{
  \noindent $\bullet$ \textbf{#1}
}

\section{Introduction}

Interrelations between the low-energy limit of supersymmetric gauge theories called Seiberg-Witten theory \cite{SW1,SW2} and $2d$ conformal theories (the AGT correspondence \cite{AGT1,AGT2,AGT3}) is one of the main issues in string theory of the last two decades. In fact, the main players behind the scene are integrable systems \cite{GKMMM1,GKMMM2,GKMMM3} and Nekrasov functions \cite{LMNS1,LMNS2,LMNS3}. Recently, it has become clear that there is the third important ingredient: the Macdonald theory \cite{Mac} and its generalizations. In particular, this is related with the algebraic setup underlying the AGT correspondence, the Ding-Iohara-Miki (DIM) algebra \cite{DI1,DI2,DI3,DI4,DI5,DI6,MMZ1} and its generalizations \cite{Saito1,Saito2,MMZ1}. In particular, the Hamiltonians of the integrable system underlying Seiberg-Witten theory are elements of the commutative subalgebra of these algebras.
Eigenfunctions of these Hamiltonians \cite{CF,MMP,MMP1,MMP2} are related with a notion of triad.

In \cite{MMP3,MMPZ1}, we defined {\bf triad} as a power series of $2N$ complex variables $x_i$, $y_i$ $i=1,\ldots,N$ that admits {\bf two polynomial reductions}: one, at $y_i=q^{\mu_i}t^{N-i}$ with $\{\mu_i\}$ being a partition, giving rise to a symmetric polynomial, another one, at $t=q^{-m}$, $m\in\mathbb{Z}_{\ge 0}$ leading to a non-symmetric (quasi)polynomial Baker-Akhiezer (BA) function\footnote{The BA function is a polynomial only up to a prefactor which becomes a polynomial at particular values of $y_i$. Hereafter, we apply the term polynomial to the quasipolynomial BA function as well.}, the latter one generating the former one at $y_i=q^{\mu_i}t^{N-i}$ with $\{\mu_i\}$ being a partition {\it only after summing} over the Weyl group. The simplest example of the triad, the basic triad presented in \cite{MMP3} is given by the Noumi-Shiraishi power series \cite{NS}, and the two polynomial reductions are just the ordinary Macdonald polynomials \cite{Mac} and the Chalykh BA function \cite{Cha}.

The basic triad admits three types of deformations. First of all, this is an extension to arbitrary root systems. Indeed, one can construct the Macdonald polynomials \cite{Macrs,Chrs1,Chrs2} and the BA function \cite{Cha} for any root system, but a counterpart of the Noumi-Shiraishi function for the case of other than $A_{N-1}$ root systems is not known. The latter still has to be eigenfunctions of the corresponding Koornwinder-Macdonald operators \cite{Macrs,Koorn}, however, explicit formulas for the Noumi-Shiraishi functions are not known in this case.

The second deformation is to the twisted case, when one knows the BA function \cite{CE,CF}, and one can construct a counterpart of the Macdonald polynomial \cite{MMP1}, but again the corresponding Noumi-Shiraishi function is not known.

At last, there is the third deformation involving elliptic deformations. We discuss in \cite{MMPZ1} the elliptic triad, which can be constructed via an elliptic deformation with an elliptic parameter $w$, and this deformation gives rise to eigenfunctions of the Hamiltonians dual to the elliptic Ruijsenaars Hamiltonians. There is another elliptic deformation with an elliptic parameter $p$, which is called the Shiraishi function \cite{Shi,LNS,AKMM1}, which, in the Nekrasov-Shatashvili limit, gives rises to polynomials that are eigenfunctions of the elliptic Ruijsenaars Hamiltonians. Hence, on the integrable theory side, the first of these deformations corresponds to the system with an elliptic dependence on momenta, while the second deformation, to the systems with an elliptic dependence on coordinates.
The system that unities both of these elliptic dependencies is called {\it bi-elliptic}, and an appropriate adjustment of Hamiltonians makes it self-dual, then it is named {\it double-elliptic} (DELL). In the present case, we deal with a {\it bi-elliptic} system, which is described by an elliptic lift of the Shiraishi function (ELS function) \cite{AKMM2}, which, in the Nekrasov-Shatashvili limit, leads to eigenfunctions \cite{MMZ} of the bi-elliptic Koroteev-Shakirov (KS) Hamiltonians \cite{KS}.

This still does not generate eigenfunctions of the DELL Hamiltonians, which were introduced in \cite{BMMM1,BMMM2,BMMM3} in order to describe $6d$ Seiberg-Witten theories with adjoint matter: in the degenerate case, the elliptic triad is related to eigenfunctions of the degenerate DELL Hamiltonians by bi-orthogonality w.r.t. the Schur scalar product \cite{MMZ}, however, this relation has not been lifted to the non-degenerate (bi-elliptic $\leftrightarrow$ DELL) case, i.e. to the ELS triad yet (see a detailed discussion of these issues in \cite{MMdell}).

Our goal in this note is to study first this second elliptic deformation, i.e. we start with construction of the triad associated with the Shiraishi function, and then we extend it to the ELS triad. One of the essential new features of the Shiraishi and ELS triads is that, under polynomial reductions, they generates infinite sets of polynomials.
Hence, we extend the notion of the triad. However, still there are two reductions: one of them gives rise to an infinite set of symmetric polynomials labeled by a partition, another one, to an infinite set of BA functions (nicknamed this way after the paper \cite{CFV}) labeled by $N$ complex parameters.
One may say that the new triads contain infinitely many subtriads each admitting two proper polynomial reductions.

The diamond of four triads looks as follows.

\bigskip

{\footnotesize
\hspace{-1cm}
\mbox{\parbox{15.6cm}
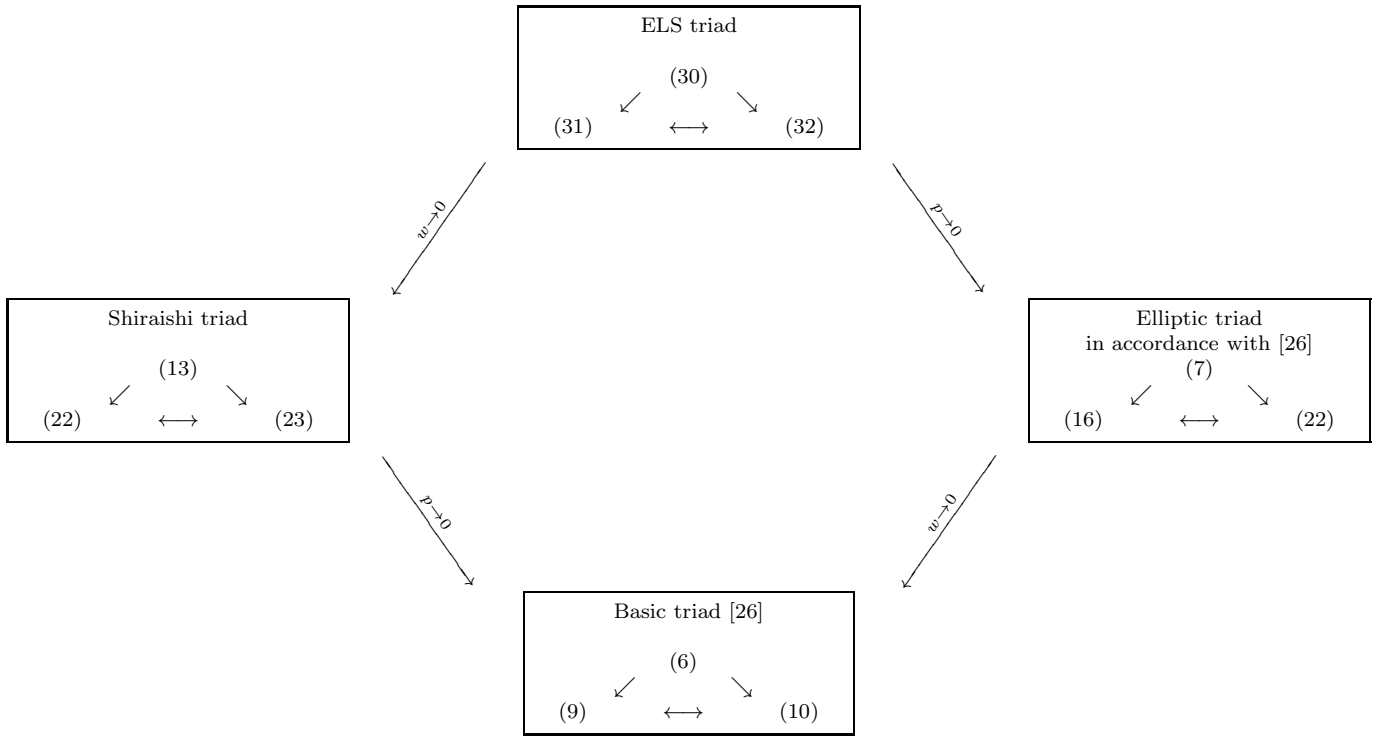
{
$$
\begin{array}{rcccl}
&&\boxed{\begin{array}{c}\hbox{ELS triad}\cr
\cr
\begin{array}{rcccl}
&&(\ref{ELS})&&\cr
&\swarrow&&\searrow&\cr
(\ref{MELS})&&\longleftrightarrow&&(\ref{BAELS})\end{array}
\end{array}}&&\cr
&{\rotatebox[origin=c]{55}{\(\xleftarrow{
{\footnotesize
\hspace*{.7cm}w\to 0\hspace*{.7cm}}
}\)}}&&{\rotatebox[origin=c]{-55}{\(\xrightarrow{\hspace*{.7cm}p\to 0\hspace*{.7cm}}\)}}&\cr
\boxed{\begin{array}{c}\hbox{Shiraishi triad}\cr
\cr
\begin{array}{rcccl}
&&(\ref{S})&&\cr
&\swarrow&&\searrow&\cr
(\ref{MS})&&\longleftrightarrow&&(\ref{BAS})\end{array}
\end{array}}
&&&&
\boxed{\begin{array}{c}\hbox{Elliptic triad}\cr
\hbox{in accordance with \cite{MMP3}}
\cr
\begin{array}{rcccl}
&&(7)&&\cr
&\swarrow&&\searrow&\cr
(16)&&\longleftrightarrow&&(22)\end{array}
\end{array}}
\cr
&{\rotatebox[origin=c]{-55}{\(\xrightarrow{\hspace*{.7cm}p\to 0\hspace*{.7cm}}\)}}&&{\rotatebox[origin=c]{55}{\(\xleftarrow{
{\footnotesize \hspace*{.7cm}w\to 0\hspace*{.7cm}}
}\)}}&\cr
&&\boxed{\begin{array}{c}\hbox{Basic triad \cite{MMP3}}\cr
\cr
\begin{array}{rcccl}
&&(\ref{NS})&&\cr
&\swarrow&&\searrow&\cr
(\ref{NSM})&&\longleftrightarrow&&(\ref{NBA})\end{array}
\end{array}}
\end{array}
$$
}}
}

\bigskip

In terms of parameters, the most general ELS function at the top of the picture depends on $2N$ complex variables $x_i$ and $y_i$, and is parameterized by two elliptic parameters, $p$ and $w$, and by three more parameters $s$, $q$ and $t$. On the physical side, this function is associated with the $6d$ Seiberg-Witten theory with adjoint matter, and these parameters have the following meaning \cite{AKMM2}:
\begin{itemize}
\item[$q$, $s$] are two $\Omega$-background deformation parameters on the gauge theory side, $s$ governs the non-stationarity. The limit of $s\to 1$ reduces the system to the quantum integrable system, and the non-stationary KS equation to the eigenvalue bi-elliptic Hamiltonian problem. On the conformal theory side, they rescale the dimensions of the operators.
\item[$t$] is the central charge parameter on the conformal theory side, and the coupling constant parameter on the integrable side. The integrable system becomes free upon $t\to 1$. On the gauge theory side, it is related to the mass $m$ of the adjoint hypermultiplet, $t=e^{-m}$.
\item[$p$] is the elliptic parameter that controls the coupling in the gauge theory (the bare torus and the bare charge). In the limit $p\to 0$, the instanton corrections disappear, and one gets the perturbative limit. On the integrable side, it is associated with the torus where the coordinates live. On the conformal theory side, it is associated with the torus where the $2d$ fields live.
\item[$w$] is the elliptic parameter that is associated with the Kaluza-Klein torus in the gauge theory (remind that one considers Seiberg-Witten $6d$ theory with two dimensions compactified onto a $2d$ torus). On the integrable side, it is associated with the torus where the momenta live.
\end{itemize}
\be
w=e^{2\pi i\hat\tau},\quad p=e^{2\pi i\tau},\quad t=e^{-m},\quad q=e^{-2\pi i\epsilon_1},\quad s=e^{-2\pi i\epsilon_2}
\ee

\paragraph{Notation.}
We use the $q$-Pochhammer symbols
\be\label{q-shift}
(u ; q)_\infty := \prod_{i=0}^\infty ( 1- q^i u), \qquad (u ; q)_n: = \frac{( u ; q)_\infty}{(q^n u ; q)_\infty}, \qquad
(u ; q,w)_\infty :=\prod_{i,j=0}^\infty ( 1- q^iw^j u)
\ee
We also define the odd theta-function as
\be
\theta_w(z):=(z;w)_\infty(w/z;w)_\infty={1\over (w;w)_\infty}\sum_{n\in\mathbb{Z}}(-1)^n
w^{n(n-1)\over 2}z^n
\ee
We also introduce the function
\be
{\cal R}(x,n):={(x;q)_n\over (xqt^{-1};q)_n}=\prod_{r=0}^{n-1}{(1-q^rx)\over(1-q^{r+1}t^{-1}x)}
\ee
and its elliptic counterpart
\be
{\cal R}_p^{ell}(x,n):=\prod_{r=0}^{n-1}{\theta_p(q^rx)\over\theta_p(q^{r+1}t^{-1}x)}
\ee

\section{Basic triad}

We start with the basic triad, which is given by the Noumi-Shiraishi power series \cite{NS}, which is a formal power series of $2N$ variables $x_i=q^{z_i}$ and $y_i=q^{\lambda_i}$, $i=1,\ldots,N$
\be\label{NS}
\boxed{
\NS_{q,t}(\vec x,\vec y)= \prod_{i=1}^Nx_i^{\lambda_i}\cdot t^{-\vec\lambda\cdot\vec\rho}\cdot
\sum_{k_{ij}}\psi_{N,N}\Big(\vec y,\{k_{ij}\};q,t\Big)\prod_{1\le i<j\le N}\left({x_j\over x_i}\right)^{k_{ij}}
}
\ee
where $\vec\rho$ is the Weyl vector, i.e. $\vec\rho\cdot\vec z={1\over 2}\sum_{i=1}^N(N-2i+1)z_i$, and the sum goes over all non-negative integer $k_{ij}$ with $i< j$, while the coefficients $\psi_{N,M}(\vec y,\{k_{ij}\};q,t)$ for any natural pair $(N,M)$: $N\le M$ are defined to be
\be\label{psiNS}
\psi_{N,M}(\vec y,\{k_{ij}\};q,t):&=&\prod_{i=1}^N\left[{\displaystyle{\prod_{{j,n}\atop{i<j\le n}}^M{\cal R}
\Big(t\cdot q^{\lambda_j-\lambda_i+\sum_{a>n}^M(k_{ia}-k_{ja})},k_{in}\Big)}
\over \displaystyle{\prod_{{j,n}\atop{i\le j<n}}^M{\cal R}\Big(q^{\lambda_j-\lambda_i-k_{jn}+\sum_{a>n}^M(k_{ia}-k_{ja})},k_{in}\Big)}}\right]
\ee

This power series enjoys the two properties: (i) it is an eigenfunction of the Macdonald-Ruijsenaars operator\cite{Mac,Rui1,Rui2,Rui3,Rui4} ({\bf Hamiltonian equation}),
\be\label{MR}
\hat H_{MR}=\sum_{i=1}^N\prod_{j\ne i}{tx_i-x_j\over x_i-x_j}q^{x_i\p_{i}}\nn\\
\hat H_{MR}\cdot\NS_{q,t}(\vec x,\vec y)=\left(\sqrt{t}\sum_iy_i\right)\cdot\NS_{q,t}(\vec x,\vec y)
\ee
and (ii) it {\bf admits two polynomial reductions}. First of all, the power series (\ref{NS}) can be made a symmetric polynomial in variables $x_i$ by choosing $\vec\lambda=\vec\mu+\vec\rho\log_q t$, where $\vec\mu$ has all non-negative integer components $\mu_j$, $\mu_1\ge\mu_2\ge\ldots\ge\mu_N\ge0$ at $j=1,\ldots,N$. This symmetric polynomial is nothing but the Macdonald polynomial \cite{NS}:
\be\label{NSM}
\boxed{
M_\mu(\{x_i\};q,t)=\NS_{q,t}(\vec x,\vec \mu+\vec\rho\log_q t)
}
\ee
Instead of getting symmetric polynomials with integrality requirements for $\vec\lambda$, one can choose $t=q^{-m}$, which gives rise to a polynomial of $x_i$ though non-symmetric, but instead $\vec\lambda$ can be kept to be $N$ arbitrary complex parameters. This polynomial is nothing but the BA function $\Psi_m(\vec x,\vec\lambda)$ \cite{Cha}:
\be\label{NBA}
\boxed{
\Psi_m(\vec x,\vec y)=\NS_{q,q^{-m}}(\vec x,\vec y)
}
\ee
which can be unambiguously (up to normalization) determined by the periodicity property giving rise to a system of linear equations:
\be\label{leq}
\Psi_m(x_kq^j,\vec y)=\Psi_m(x_lq^j,\vec y)\ \ \ \ \  \forall k,l\ \ \hbox{and}\ \ 1\le j\le m\ \ \ \ \ \hbox{at}\ \ x_k=x_l
\ee
We keep this term in all other cases: we call BA function the reduction at $t=q^{-m}$.

In our previous paper \cite{MMP3,MMPZ1}, we considered the basic and elliptic triads, which are based on power series that admit two polynomial reductions: one of them gives rise to the BA function defined at $N$ arbitrary complex parameters $\vec\lambda$, the other one leads to a symmetric polynomial when $\vec\lambda$ is a partition, and can be obtained from the BA function by summation over the Weyl group.

In this letter, we use the method due to \cite{LNS} of lifting the Noumi-Shiraihsi function to the Shiraishi function (see sec.3.1) in order to lift the elliptic generalization of the Noumi-Shiraishi function \cite{AKMM2} to the ELS function \cite{AKMM2} (see sec.4). This is done much similar to the line of \cite{LNS}.
We also observe that, in the both Shiraishi and ELS cases, each of the two polynomial reductions gives rise to an infinite set of polynomials such that the BA function is a set of polynomials defined at $N$ arbitrary complex parameters $\vec\lambda$, and the second set of polynomials are symmetric polynomials when $\vec\lambda$ is a partition. Symmetric polynomials from the second set are obtained from polynomials of the first (BA) set by summation over the Weyl group. We checked this fact with the computer for many particular cases. Various symmetric properties of the triads and linear equations similar to (\ref{leq}) in the Noumi-Shiraishi case are also discussed.

Thus, new are formula (\ref{ELS}) with (\ref{psiELS}) for the ELS function \cite{AKMM2} and the relations of two polynomial reductions (\ref{MPsi})-(\ref{eq}) both in the cases of Shiraishi and ELS triads.

\section{Shiraishi triad}

\subsection{The Shiraishi function}

\paragraph{Definition.} In order to introduce an elliptic parameter $p$, one can achieve a periodicity by choosing
\be\label{pc}
x_{i+N}=px_i,\ \ \ \ \ \ \ \ y_{i+N}=sy_i
\ee
in the system with infinitely many $x_i$ and $y_i$. This procedure, however, requires some regularization. One way to make it is to restrict
all the products over $i$ by $N$ \cite{LNS} and keep all other products up to infinity so that
(\ref{NS}) changes for
\be\label{S}
\boxed{
\S_N(\vec x ; p \vert \vec y ; s\vert q,t)= \prod_{i=1}^Nx_i^{\lambda_i}\cdot t^{-\vec\lambda\cdot\vec\rho}\cdot
\sum_{k_{ij}}\psi_{N,\infty}\Big(\vec y,\{k_{ij}\};q,t,s\Big)\prod_{i=1}^N\prod_{j>i}^\infty\left({x_j\over x_i}\right)^{k_{ij}}
}
\ee
with the periodic conditions imposed on $k_{i+N,j+N}=k_{ij}$, and (\ref{pc}) for $x_i$, $y_i$ .

$\S_N(\vec x ; p \vert \vec y ; s\vert q,t)$ is just the Shiraishi function \cite{Shi,LNS}.

\subsection{Properties of the Shiraishi function}

\paragraph{Bispectral duality.} Upon choosing a proper normalization \cite[Formula (1)]{Shi},
\be
\S_N^{(1)}(\vec x ; p \vert \vec y ; s\vert q,t):= {\cal N}_1(\vec y,s\vert q,t)\S_N(\vec x ; p \vert \vec y ; s\vert q,t)\nn\\
{\cal N}_1(\vec y,s\vert q,t):=\prod_{i<j}{\Big(q{y_j\over y_i};q,s^N\Big)_{\!\infty}\over \Big({qy_j\over ty_i};q,s^N\Big)_{\!\infty}}\cdot
\prod_{i\le j}{\Big(qs^N{y_i\over y_j};q,s^N\Big)_{\!\infty}\over \Big(s^N{qy_i\over ty_j};q,s^N\Big)_{\!\infty}}
\ee
the Shiraishi function becomes symmetric under permuting the sets $(\vec x,p)\longleftrightarrow (\vec y,s)$ \cite[Formula (2)]{Shi}, \cite[Formula (2.12)]{LNS},
\be\label{symm1}
\S_N^{(1)}(\vec x ; p \vert \vec y ; s\vert q,t)=\S_N^{(1)}(\vec y ; s\vert \vec x ;p\vert q,t)
\ee
and so does the Noumi-Shiraishi function (\ref{NS}) \cite[Formula (1.17)]{NS} \cite{Cha,MMP1} with a similar normalization factor
\be
\widehat{\cal N}_1(\vec y\vert q,t):=\prod_{i<j}{\Big(q{y_j\over y_i};q\Big)_\infty\over \Big({qy_j\over ty_i};q\Big)_\infty}
\ee

\paragraph{Poincar\'e duality.}
Further, choosing another normalization \cite[Formula (1)]{Shi},
\be
\S_N^{(2)}(\vec x ; p \vert \vec y ; s\vert q,t):= {\cal N}_2(\vec y,s\vert q,t)\S_N(\vec x ; p \vert \vec y ; s\vert q,t)\nn\\
{\cal N}_2(\vec y,s\vert q,t):=\prod_{i<j}{\Big({qx_j\over tx_i};q,p^N\Big)_\infty\over \Big({qx_j\over x_i};q,p^N\Big)_\infty}\cdot
\prod_{i\le j}{\Big(p^N{qx_i\over tx_j};q,p^N\Big)_\infty\over \Big(p^N{qx_i\over x_j};q,p^N\Big)_\infty}
\ee
the Shiraishi function does not change under the transformation $t\to q/t$ \cite[Formula (3)]{Shi}, \cite[Formula (2.12)]{LNS},
\be\label{symm2}
\S_N^{(2)}(\vec x ; p \vert \vec y ; s\vert q,t)=\S_N^{(2)}(\vec x ; p \vert \vec y ; s\vert q,q/t)
\ee
and so does the Noumi-Shiraishi function (\ref{NS}) \cite[Formula (1.17)]{NS} \cite{Cha,MMP1} with the normalization factor
\be
\widehat{\cal N}_2(\vec y\vert q,t):=\prod_{i<j}{\Big({qx_j\over tx_i};q\Big)_\infty\over \Big({qx_j\over x_i};q\Big)_\infty}
\ee

\paragraph{Hamiltonian equation.} The equation for the Shiraishi function is rather involved \cite{LNS}, and is nicknamed ``non-stationary elliptic Ruijsenaars equation" for the following reason: in the limit of\footnote{In this limit, the Shiraishi function is singular, and one has to change the normalization, so that the Shiraishi function becomes regular in the limit of $s\to 1$.} $s\to 1$, the Shiraishi function becomes an eigenfunction of the elliptic Ruijsenaars Hamiltonian \cite{Shi},
\be\label{MRell}
\hat H_{ellR}&=&\sum_{i=1}^N\prod_{j\ne i}{\theta_p(tx_i/x_j)\over \theta_p(x_i/x_j)}q^{x_i\p_{i}}\nn\\
\hat H_{ellR}\cdot\S_N(\vec x ; p \vert \vec y ; s\to 1\vert q,t)&=&\Lambda(y_i;p,q,t)\cdot\S_N(\vec x ; p \vert \vec y ; s\to 1\vert q,t)
\ee
Thus, in accordance with what we explained in the Introduction, $s$ controls the non-stationarity, and the limit of $s\to 1$ corresponds to the Nekrasov-Shatashvili limit.

\subsection{Polynomial reductions}

In order to generate polynomial reductions of the Shiraishi function, one has to expand it in powers of $p$:
\be
\S_N(\vec x ; p \vert \vec y ; s\vert q,t)=\sum_{n\ge 0}p^{Nn}\cdot{\S_N^{(n)}(\vec x ; p \vert \vec y ; s\vert q,t)\over \prod_{i=1}^N x_i^n}=
\sum_{n\ge 0}\S_N^{(n)}(\vec x ; p \vert \vec y ; s\vert q,t)\cdot\prod_{i=1}^N\left({p\over x_i}\right)^n
\ee

\paragraph{Reduction to symmetric polynomials.}
Thus, one gets an infinite set of power series $\S_N^{(n)}(\vec x ; p \vert \vec y ; s\vert q,t)$ labeled by an additional {\it level} $n$,
each of them admitting two polynomial reductions. The first one emerges upon choosing
$y_i=q^{\mu_i}t^{N-i}$ with $\mu_i\in\mathbb{Z}$, $\mu_1\ge\mu_2\ge\ldots\mu_N\ge 0$, and multiplying $\S_N^{(n)}(\vec x ; p \vert \vec y ; s\vert q,t)$ with $\prod_{i+1}^Nx_i^{\mu_i}$:
\be\label{MS}
\boxed{
\mathfrak{M}_\mu(\vec x ; p; s\vert q,t)=\S_N\Big(\vec x ; p \big| \{q^{\mu_i}t^{N-i}\} ; s\big| q,t\Big)=
\sum_{n\ge 0}\mathfrak{M}_\mu^{(n)}(\vec x ; s\vert q,t)\cdot\prod_{i=1}^N\left({p\over x_i}\right)^n
}
\ee
$\mathfrak{M}_\mu^{(n)}(\vec x ; s\vert q,t)$ is a symmetric polynomial, and $\mathfrak{M}_\mu^{(0)}(\vec x ; s\vert q,t)$ does not depend on $s$, and is the standard Macdonald polynomial $M_\mu(\vec x;q,t)$.

\paragraph{BA reduction.} Another polynomial reduction emerges upon choosing $t=q^{-m}$, $m\in\mathbb{Z}_{\ge 0}$.
This polynomial reduction gives rise to the BA function:
\be\label{BAS}
\boxed{
\Psi_m(\vec x;p\,|\,\vec y;s)=
\S_N(\vec x ; p \vert \vec y ; s\vert q,q^{-m}):=\sum_n\Psi_m^{(n)}(\vec x;\vec y;s)\prod_{i=1}^N\left({p\over x_i}\right)^n
}
\ee
 and
\be\label{BAS2}
\boxed{
\Psi_m^{(n)}(\vec x;\vec y;s)=\prod_{i=1}^Nx_i^{\lambda_i-m\rho_i}\cdot\sum_{k_{ij}}
\psi_m^{(n)}\Big(\vec y,\{k_{ij}\},s;q\Big)\prod_{1\le i<j\le N}\left({x_j\over x_i}\right)^{k_{ij}}
}
\ee
with $0\le k_{ij}\le m+2n$,
and $\Psi_m^{(n)}(\vec x;\,|\,\vec y;s)$ still satisfies the same periodicity property (\ref{leq}):
\be
\Psi_m(x_kq^j,\vec y)=\Psi_m(x_lq^j,\vec y)\ \ \ \ \  \forall k,l\ \ \hbox{and}\ \ 1\le j\le m\ \ \ \ \ \hbox{at}\ \ x_k=x_l
\ee
In fact, $\Psi_m^{(0)}(\vec x;\,|\,\vec y;s)=\Psi_m^{(0)}(\vec x;\,|\,\vec y)$ does not depend on $s$.

Two remarks are in order. First of all, the both symmetricity properties (\ref{symm1}) and (\ref{symm2}) are inherited by the full BA function $\Psi_m(\vec x;p\,|\,\vec y;s)$, but not by the separate polynomials $\Psi_m^{(n)}(\vec x;\vec y;s)$ at $n\ne 0$. In particular, this means that, in contrast with the $n=0$ case, one cannot use the BA function for constructing symmetric polynomials at $t=q^{m+1}$.

Second, note that the periodicity property is enough to unambiguously fix the BA function $\Psi_m^{(0)}(\vec x;\,|\,\vec y)$ \cite{Cha}, however, this is no longer the case for higher $\Psi_m^{(n)}(\vec x;\,|\,\vec y;s)$: the number of equations necessary for fixing the BA function $\Psi_m^{(n)}(\vec x;\,|\,\vec y;s)$ grows with $n$. It is, therefore, an open question
whether periodicity equations at $n\neq0$ can be augmented by more equations
so that each $\Psi_m^{(n)}(\vec x;\,|\,\vec y;s)$ could be defined purely from these
equations unambiguously.

\paragraph{Relation of the two reductions.}
In the case of the basic triad, one can make the Macdonald polynomial from the BA function at non-negative integer components of $\vec\lambda+m\vec\rho$ ordered in non-increasing order (and this is Chalykh's original relation between the two, see \cite[Theorem 5.11]{Cha}), with help of averaging over Weyl group action:
\be\label{MPsi1}
M_{\vec\lambda+m\vec\rho}(\vec x;q,q^{-m})=\sum_{w\in W}\Psi_m(w\vec x,\vec\lambda)
\ee
Here $w$ is an element of the $A_{N-1}$ Weyl group. It turns out that this relation persists to be correct at the level of the Shiraishi functions:
\be\label{MPsi}
\mathfrak{M}_{\vec\lambda+m\vec\rho}^{(n)}(\vec x ; p; s\vert q,q^{-m})=\sum_{w\in W}\Psi_m^{(n)}(w\vec x;\vec y;s)
\ee
or, equivalently,
\be\label{eq}
\S_N(\vec x ; p \vert \{q^{\mu_i}t^{N-i} ; s\vert q,t)=\sum_{w\in W}\S_N(w\vec x ; p \vert \vec y ; s\vert q,q^{-m})
\ee

\section{ELS triad}

\paragraph{Definition.}
Now, this is absolutely immediate to construct the ELS triad: one just extends the coefficients in (\ref{psiNS}) replacing differences for the theta-functions as it was done in \cite{FOS,AKMM2,MMPZ1}:
\be\label{psiELS}
\psi_{N,M}^{ell}\Big(\vec y,\{k_{ij}\};q,t,w\Big):&=&\prod_{i=1}^N\left[{\displaystyle{\prod_{{j,n}\atop{i<j\le n}}^M{\cal R}_w^{ell}
\Big(t\cdot q^{\lambda_j-\lambda_i+\sum_{a>n}^M(k_{ia}-k_{ja})},k_{in}\Big)}
\over \displaystyle{\prod_{{j,n}\atop{i\le j<n}}^M{\cal R}^{ell}_w\Big(q^{\lambda_j-\lambda_i-k_{jn}+\sum_{a>n}^M(k_{ia}-k_{ja})},k_{in}\Big)}}\right]
\ee
Then, again implying the periodic conditions imposed on $k_{i+N,j+N}=k_{ij}$, and (\ref{pc}) for $x_i$, $y_i$, one immediately obtains the ELS power series \cite{AKMM2}
\be\label{ELS}
\boxed{
\S_N^{ell}(\vec x ; p \vert \vec y ; s\vert q,t,w)= \prod_{i=1}^Nx_i^{\lambda_i}\cdot t^{-\vec\lambda\cdot\vec\rho}\cdot
\sum_{k_{ij}}\psi_{N,\infty}^{ell}\Big(\vec y,\{k_{ij}\};q,t,w\Big)\prod_{i=1}^N\prod_{j>i}^\infty\left({x_j\over x_i}\right)^{k_{ij}}
}
\ee

\paragraph{Polynomial reductions.}
In this case, one literally repeats the steps we did in the case of the Shiraishi triad, in order to get the infinite set of symmetric polynomials
\be\label{MELS}
\boxed{
\mathfrak{M}_\mu^{ell}(\vec x ; p; s\vert q,t,w)=\S_N\Big(\vec x ; p \big| \{q^{\mu_i}t^{N-i}\} ; s\big| q,t,w\Big)=
\sum_{n\ge 0}\mathfrak{M}_\mu^{(ell,n)}(\vec x ; s\vert q,t,w)\cdot\prod_{i=1}^N\left({p\over x_i}\right)^n
}
\ee
and the infinite set of the BA functions:
\be\label{BAELS}
\boxed{
\Psi_m^{ell}(\vec x;p\,|\,\vec y;s\vert q,w)=
\S_N^{ell}\Big(\big| x ; p \vert \vec y ; s\big| q,q^{-m},w\Big):=\sum_n\Psi_m^{(ell,n)}(\vec x;\vec y;s,w)\prod_{i=1}^N\left({p\over x_i}\right)^n
}
\ee
Note that the fact that $\mathfrak{M}_\mu^{(ell,n)}(\vec x ; s\vert q,t,w)$ are symmetric polynomials is highly non-trivial \cite{AKMM3}, and is a corollary of a series of theta-function identities \cite{AKMM2}. The relation between the two polynomial reductions, (\ref{MPsi})-(\ref{eq}) still persists in this ELS triad case. However, a counterpart of (\ref{leq}) in the elliptic case is yet to be constructed \cite{MMPZ1}.

\paragraph{Hamiltonian equation.} Similarly to the Shiraishi triad case, in order to get a ``stationary" equation in the ELS triad case, one has to take the limit of $s\to 1$ with a proper changing of normalization. Then, the ELS triad becomes an eigenfunction of the KS Hamiltonians \cite{MMZ}. These Hamiltonians can be presented either \cite[Eqs.(97),(137)]{MMdell} in the original KS form \cite{KS}, or \cite[Eqs.(127),(138)]{MMdell} in the form proposed in \cite{GZ}. The Hamiltonians are non-local and quite involved, and we do not reproduce them here.

\section{Conclusion}

In this letter, we considered an additional elliptic deformation with an elliptic parameter $p$ of known triads:
the basic triad from \cite{MMP3} and the elliptic triad from \cite{MMPZ1}.
This allowed us to extend the notion of triad so that it allows both reductions to infinite sets of polynomials:
(symmetric) $\mathfrak{M}_\mu^{(n)}(\vec x ; s\vert q,t)$
and
$\Psi_m^{(n)}(\vec x;\,|\,\vec y;s)$ ($n=0,\ldots,\infty$) obtained by an expansion of the Shiraishi function in $p$,
(symmetric) $\mathfrak{M}_\mu^{(ell,n)}(\vec x ; p; s\vert q,t,w)$ and
$\Psi_m^{(ell,n)}(\vec x;p\,|\,\vec y;s\vert q,w)$ ($n=0,\ldots,\infty$) obtained by an expansion of the ELS function in $p$.
The resulting {\bf bi-elliptic ELS triad} is the largest triad associated with $6d$ Seiberg-Witten theory with adjoint matter,
and the four constructed triads exhaust polynomial systems associated with Seiberg-Witten theories of this type.
An open question remains of bringing it to an explicit DELL \cite{MMdell} (self-dual) form.

\section*{Acknowledgements}

This work was supported by the Russian Science Foundation (Grant No.23-41-00049).

\end{document}